\documentclass[prl,twocolumn]{revtex4}
\usepackage{graphicx, epsfig}
\usepackage{color}
\usepackage{mathrsfs}
\usepackage{bm}
\usepackage{amsmath}

\newcommand{\be}{\begin{equation}}
\newcommand{\ee}{\end{equation}}
\newcommand{\bea}{\begin{eqnarray}}
\newcommand{\eea}{\end{eqnarray}}
\newcommand{\ba}{\begin{eqnarray}}
\newcommand{\ea}{\end{eqnarray}}

\newcommand{\gapp}{\mathrel{\raise.3ex\hbox{$>$}\mkern-14mu
              \lower0.6ex\hbox{$\sim$}}}
\newcommand{\lapp}{\mathrel{\raise.3ex\hbox{$<$}\mkern-14mu
              \lower0.6ex\hbox{$\sim$}}}

\begin{document}
\title{Fundamental Implications of Intergalactic Magnetic Field Observations}
\author{Tanmay Vachaspati}
\affiliation{
Physics Department, Arizona State University, Tempe, AZ 85287, USA.
}

\begin{abstract}
\noindent
Helical intergalactic magnetic fields at the $\sim 10^{-14}~{\rm G}$ level on $\sim 10~{\rm Mpc}$ length 
scales are indicated by current gamma ray observations. The existence of magnetic fields in cosmic voids
and their non-trivial helicity suggest that they must have originated in the early universe and thus have 
implications for the fundamental interactions. I combine present knowledge of the observational
constraints and the dynamics of cosmological magnetic fields to derive characteristics that would
need to be explained by the magnetic field generation mechanism.
The importance of CP violation and a possible crucial role for chiral effects in the early universe are pointed out.
\end{abstract}

\maketitle

Several independent investigations of gamma rays from blazars indicate the presence of intergalactic 
magnetic fields \cite{Neronov:1900zz,Ando:2010rb,Essey:2010nd,Tashiro:2013ita,Chen:2014qva,
Chen:2014rsa}. 
Emission of TeV energy gamma rays from blazars and the subsequent electromagnetic cascade
in the intergalactic medium is expected to distort the intrinsic blazar spectrum by depleting photons
from the TeV range and adding photons in the GeV range. The lack of expected additional photons in the
GeV range is explained by invoking an intergalactic magnetic field of strength $\gtrsim 10^{-16}~{\rm GeV}$.
As an intergalactic magnetic field disperses the additional GeV photons, the intergalactic magnetic field
hypothesis also predicts a halo of GeV photons around the blazar. An analysis of stacked blazars provides
evidence for such a halo and adds support to the derived lower bound on intergalactic magnetic 
fields \cite{Chen:2014rsa}.

An alternative approach developed in Refs.~\cite{Tashiro:2013bxa,Tashiro:2014gfa} utilizes the {\it helical} 
nature of intergalactic magnetic fields. The reasoning is that intergalactic magnetic fields are measured 
in cosmic voids, $\sim 100~{\rm Mpc}$ away from astrophysical sources, and thus were most likely 
generated in the early universe. (For a review of magnetic fields and some possible astrophysical 
generation mechanisms see Ref.~\cite{Durrer:2013pga}.)
 Unless the magnetic fields are coherent on very long length
scales or are helical at the time of production, they would dissipate and not survive until the present epoch. 
If the magnetic field generation mechanism was causal, the magnetic fields are not coherent on large length scales 
and helicity is essential for survival. Furthermore, the observation of helicity can help distinguish between 
cosmological and astrophysical magnetic fields as a globally preferred sign of the helicity would be indicative 
of a fundamental production mechanism.

In Refs.~\cite{Tashiro:2013bxa,Tashiro:2014gfa} it was shown that the helicity of the intergalactic 
magnetic field leaves a parity odd imprint on the distribution of cascade gamma rays. Thus helicity 
can be deduced by calculating parity odd correlators of observed gamma ray arrival directions. 
(Simulations of the process can be found in \cite{Long:2015bda,Batistaetal}.)
Using this technique, it becomes possible to {\it measure} --
not jut bound -- the power spectra of intergalactic magnetic fields. Applying this technique on 
current Fermi-LAT data, Refs.~\cite{Tashiro:2013ita,Chen:2014qva}
estimate the intergalactic magnetic field to be $\sim 10^{-14}~{\rm G}$ as measured on a length 
scale $\sim 10~{\rm Mpc}$ . The statistical significance of these measurements is at $\sim 3.5\sigma$ 
level in analysis with current data \cite{FerrerVachaspati16}. Further observations, especially using
a variety of observational tools, will be able to confirm or refute these
findings. For this paper we proceed on the assumption that the accumulating observational evidence 
is correct.

The existence of helical intergalactic magnetic fields points to an early universe
origin and therefore is of interest to particle cosmology. As observational dataset gets larger, 
it will become possible to measure the magnetic field correlation functions over a range of
scales. If the spectrum is flat or red, {\it i.e.} does not fall off at large length scales, the
magnetic field would likely be a product of the big bang or inflation. In this case, the
primordial magnetic field may shed light on cosmological initial conditions and it may also have 
important consequences for the origin of the matter-antimatter asymmetry 
\cite{Anber:2015yca,Fujita:2016igl} and other theoretical ideas \cite{Long:2015cza}.
If the spectrum is measured to be blue, we expect the magnetic field 
to have been produced in high energy particle processes, and the helicity of the magnetic
field points to an important role for CP violating interactions in the early universe. 

For the rest of our discussion, we will assume that the intergalactic magnetic field is stochastic
and isotropic, and is generated by a causal mechanism. (If the generation mechanism were
acausal, the field may not even be stochastic within our cosmic horizon.) Then the spatial correlation 
function of the magnetic field is given by \cite{MoninYaglom}
\begin{equation}
\langle B_i ({\bm x}) B_j ({\bm x}+{\bm r}) \rangle = M_N(r) P_{ij} + M_L(r) {\hat r}_i {\hat r}_j
+ \epsilon_{ijk} {\hat r}_k M_H (r)
\label{Bxcorrel}
\end{equation}
where $P_{ij} = \delta_{ij}-{\hat r}_i {\hat r}_j$. $M_N(r)$ and $M_L(r)$ are the ``normal'' and
``longitudinal'' power spectra and are related by a differential equation \cite{MoninYaglom};
$M_H(r)$ is the helical power spectrum and is what is measured by the parity odd gamma ray 
correlators.

Our first task is to relate the spatial helical correlation function to its counterpart in Fourier
space because the magneto-hydrodynamic (MHD) evolution of the magnetic field is
carried out in Fourier space while the field correlations are measured in physical space.
The Fourier space correlation functions for a stochastic, isotropic magnetic field are
written as
\begin{eqnarray}
\langle b_i ({\bm k}) b_j ^* ({\bm k}') \rangle &=& 
\left [ \frac{E_M(k)}{4\pi k^2} p_{ij} + i \epsilon_{ijl} k_l \frac{H_M(k)}{8\pi k^2} \right ] \nonumber \\
&&
\hskip 0.5 in \times (2\pi)^6 \delta^{(3)} ( {\bm k} - {\bm k}' )
\label{bkcorrel2}
\end{eqnarray}
where $p_{ij} = \delta_{ij}-{\hat k}_i {\hat k}_j$ and
\begin{equation}
{\bm b}({\bm k}) = \int d^3x {\bm B}({\bm x}) e^{i {\bm k}\cdot {\bm x}}, \ \ 
{\bm B}({\bm x}) = \int \frac{d^3k}{(2\pi)^3} {\bm b}({\bm k}) e^{-i {\bm k}\cdot {\bm x}}
\label{Bxbk}
\end{equation}
We now use Eq.~(\ref{Bxbk}) in (\ref{Bxcorrel}) to obtain
\begin{equation}
M_H(r) =  \frac{1}{2} \int_0^\infty dk ~k H_M(k) \frac{d}{d\rho} \left ( \frac{\sin\rho}{\rho} \right )
\label{MHHMk}
\end{equation}
where $\rho = k r$.

Studies of the MHD equations show that a cosmological magnetic field with helicity
evolves so that at late times \cite{Jedamzik:2010cy,Kahniashvili:2012uj,Saveliev:2013uva,ewBevoln}
\begin{equation}
E_M (k) = \frac{k}{2} | H_M(k) | =
\begin{cases}
E_0 (k/k_d)^4, & 0\le k \le k_d \\
0,                & k_d < k
\end{cases}
\label{EMspectrum}
\end{equation}
where the first equality is the relation for maximal helicity, the functional dependence $k^4$ defines 
the ``Batchelor spectrum'', and $k_d$ is a dissipation scale that will be discussed below. For
$k > k_d$, the spectrum falls off rapidly and so we have set it to zero. Strictly, the Batchelor
spectrum only applies for $k < k_I$ where $k_I < k_d$ is the ``inertial scale'' where the spectrum
peaks. For $k_I < k < k_d$, the spectrum falls off as a power law and there is a sharper fall off for 
$k > k_d$~\footnote{I thank Tina Kahniashvili for emphasizing this point.} \cite{Kahniashvili:2009qi}.
For simplicity, we have taken $k_I \approx k_d$, which may also be justified if the
magnetic field is generated on very small scales.
We shall also assume $H_M(k) \ge 0$ to be concrete.
Below we will estimate the power spectrum amplitude, $E_0$ in Eq.~(\ref{EMspectrum}).

Gamma ray observations have been used to measure $M_H(r)$. So we use Eq.~(\ref{EMspectrum})
in (\ref{MHHMk}) to obtain $M_H(r)$
\begin{equation}
M_H(r) = \frac{E_0 k_d}{\rho_d^5} [ (\rho_d^3-8\rho_d) \sin\rho_d + 4(\rho_d^2-2)\cos\rho_d + 8]
\label{MHr}
\end{equation}
where $\rho_d \equiv k_d r$.
One can check: $M_H(r) \propto r$ as $r\to 0$ and $M_H(r) \to \sin(k_d r)/r^2$ as $r \to \infty$,
so $M_H(r)$ is well-behaved for all $r$.

Any observation will measure a ``smeared'' $M_H(r)$. For example, gamma ray observations in 
Refs.~\cite{Tashiro:2013ita,Chen:2014qva}
measure $M_H$ on a certain distance scale $r$ that is determined from the energies of observed gamma rays.
However, for statistical purposes, the observed gamma rays are binned according to their energies
-- in 10~GeV wide bins in Refs.~\cite{Tashiro:2013ita,Chen:2014qva}. This means that observations yield 
$M_H$ that is smeared 
over a range, $\Delta r$, of $r$. With present day observations, $r$ is typically 
on the order of Mpc, and $l_d= 2\pi/k_d$ is typically kpc, so that $\rho_d = k_d r \gg 1$. The precise smearing
function depends on the binning procedure and experimental details ({\it e.g.} energy dependence of time 
exposure of the experiment), however, with current parameters $l_d \ll \Delta r \lesssim r$.

Let us write $\Delta\rho_d = k_d\Delta r$. Then, from Eq.~(\ref{MHr}), the smearing procedure will
effectively replace the oscillating trigonometric functions by (weighted) averages. For example,
\begin{equation}
\frac{\sin\rho_d}{\rho_d^2} \to \frac{1}{\Delta\rho_d} 
\int_{\rho_d}^{\rho_d+\Delta\rho_d} d\rho \frac{\sin\rho}{\rho^2}
  \approx \frac{\mathcal{O}(1)}{\rho_d^2}.
\end{equation}
Since $\rho_d \gg 1$, the 
$\rho_d^3$ term in the square bracket in Eq.~(\ref{MHr}) will dominate and we can write
\begin{equation}
M_H(r) \approx \frac{E_0 k_d}{\rho_d^2}
\end{equation}
Therefore a measurement of $M_H(r)$ at $r=r_*$, denoted $M_{H*}$, will give
\begin{equation}
E_0 = \frac{\rho_*^2 M_{H*}}{k_d}
\end{equation}
where $\rho_* = k_d r_*$, and the magnetic field energy and helicity spectra 
in Eq.~(\ref{EMspectrum}) become,
\begin{equation}
E_M(k)= \frac{k}{2} |H_M(k)| = \rho_* r_* |M_{H*}| \left ( \frac{k}{k_d} \right )^4
\end{equation}

From Eq.~(50) of Ref.~\cite{Tashiro:2014gfa}~\footnote{The definition of $M_H$ we 
are using in Eq.~(\ref{Bxcorrel}) differs from that in \cite{Tashiro:2014gfa} by a factor of $r$.} 
we have the estimate
\begin{equation}
|M_{H*}| \sim (10^{-14}~{\rm G})^2
\label{MHestimate}
\end{equation}
and $r_* \sim 10~{\rm Mpc}$. 
Subsequent (and ongoing) analyses \cite{Chen:2014qva} show rough agreement with these 
estimates and future observations should be able to pin down the values more accurately. Other analyses 
\cite{Neronov:1900zz,Ando:2010rb,Essey:2010nd,Chen:2014rsa} do not provide measurements 
of the field strength but they do provide lower bounds if they assume a coherence scale and
a spectrum. These lower bounds on the field strength 
are on the order of $10^{-16}~{\rm G}$ (see Fig.~12 of Ref.~\cite{Durrer:2013pga}). 

The energy density in the magnetic field is 
\begin{equation}
{\cal E} =  \frac{1}{2} \langle {\bm B}^2 \rangle
= \int dk ~E_M(k) \sim (10^{-14}~{\rm G})^2 \frac{\rho_*^2}{5}
\label{calE}
\end{equation}
Similarly the helicity density is given by
\begin{eqnarray}
H &=& \lim_{V \to \infty} \left \langle \frac{1}{V} \int_V d^3x {\bm A}\cdot {\bm B} \right \rangle  \nonumber \\
&=& \int dk ~H_M(k) \sim (10^{-14}~{\rm G})^2 \frac{\rho_*^2}{2k_d}
\label{helicity}
\end{eqnarray}
where ${\bm B}={\rm curl}({\bm A})$. 

Next we discuss the dissipation length scale $l_d$.
In Ref.~\cite{Jedamzik:1996wp}, the authors considered a homogeneous magnetic field and 
calculated the damping rate of small perturbations on this background. The dominant dissipation 
of the small perturbations is due to the damping of fast magnetosonic modes. Hence this mechanism 
sets the dissipation scale that then depends on the strength of the background uniform field. 

The damping of a {\em stochastic, helical} magnetic field has been discussed in
Ref.~\cite{Banerjee:2004df,Kahniashvili:2010wm}. The evolution of the dissipation scale, which
roughly coincides with the coherence scale for the Batchelor spectrum, depends 
on properties of the magnetic field at the time it was generated. The result for the
dissipation scale at the present epoch is (see Eqs.~(4) and (5) of \cite{Banerjee:2003xk})
\begin{eqnarray}
l_{d0} &=& 0.45~{\rm pc} \sqrt{n} \left ( \frac{\Omega_{B{\rm Rad} g}}{0.083} \right )^{1/2} x^{-2/(n+2)} \nonumber \\
             && \hskip 0.5 in  \times \left ( \frac{T_g}{100~{\rm MeV}} \right )^{-n/(n+2)}
\end{eqnarray}
where $n$ is the spectral index for the magnetic field, $\Omega_{B{\rm Rad} g}$ is the
ratio of the energy density in magnetic fields to that in radiation (in all relativistic species), 
$T_g$ is the temperature, and all quantities are taken at the time of magnetic field generation
(denoted by subscript ``$g$''). 
Also, $x=2.3\times 10^{-9}$ is a numerical factor. This formula yields
\begin{equation}
l_{d0} \approx 1 ~ {\rm pc} - 1~ {\rm kpc}
\end{equation}
for magnetic field generation at the electroweak epoch ($T_g =100~{\rm GeV}$), for $n=2-5$ --
larger $n$ gives smaller $l_{d0}$ --  and with 
$\Omega_{B{\rm Rad} g}=0.083$. The index $n$ is defined in \cite{Banerjee:2003xk} by the
relation $\rho_B \propto l^{-n}$ where $\rho_B$ is the energy density in magnetic fields on
a length scale $l$ {\it at the epoch of magnetogenesis}. Translating this into our language with 
the relation in Eq.~(\ref{calE}) we have $n=5$ for the Batchelor spectrum, and $n=3$ based
on a model of processes that might have occured during a first order phase transition \cite{Ng:2010mt}.


With $l_{d0} = 1~{\rm kpc}$ and $r_* = 10~{\rm Mpc}$
we get $\rho_* = 2\pi \times 10^4$. Inserting this
estimate of $\rho_*$ in Eq.~(\ref{calE}) gives the magnetic field energy density
at the present epoch,
\begin{equation}
{\cal E}_0 \sim (3\times 10^{-10} ~{\rm G})^2 ~ \left (\frac{1~{\rm kpc}}{l_{d0}} \right )^2
\label{EGauss}
\end{equation}
and Eq.~(\ref{helicity}) gives
\begin{equation}
H_0 \sim 3\times 10^{-20} ~{\rm G}^2 -{\rm kpc} ~ \left (\frac{1~{\rm kpc}}{l_{d0}} \right )
\end{equation}
In natural units ($\hbar=1=c$), with the conversions 
$1~{\rm G}= 1.95\times 10^{-20}~{\rm GeV}^2=5\times 10^7~{\rm cm}^{-2}$,
we can also write
\begin{equation}
H_0 \sim 2\times 10^{17} ~{\rm cm}^{-3}~ \left (\frac{1~{\rm kpc}}{l_{d0}} \right )
\end{equation}

To get a feel for these estimates, we compare the energy density in magnetic fields to that in
photons,
\begin{equation}
\Omega_{B\gamma0} = \frac{{\cal E}_0}{\rho_{\gamma0}} \sim 10^{-8} ~ \left (\frac{1~{\rm kpc}}{l_{d0}} \right )^2.
\label{OmegaBg}
\end{equation}
where $\rho_{\gamma0}=4.6\times 10^{-34}~{\rm gms/cm^3} \approx (4\times 10^{-6}~{\rm G})^2$ is the 
energy density in photons at the present epoch.

To proceed further we would like to estimate $\Omega_{B\gamma}$ at earlier times. The full details
of the evolution are complicated because of episodes ({\it e.g.} $e^+e^-$ annihilation), viscosity,
finite electrical conductivity, and unknown factors ({\it e.g.} neutrino masses). However 
a simple approximate picture emerges from various studies within the context of conventional MHD
Refs.~\cite{Banerjee:2004df,Kahniashvili:2012uj}. Most crucially, helicity is found to be
conserved, so the helicity density $H \propto a^{-3}$ where $a(t)$ is the cosmic scale factor. The 
dissipation scale, also the scale where
most of the magnetic energy is stored, grows as $l_d \propto a \times a^{2/3}$ in the radiation
dominated era and as $l_d \propto a$ in the matter dominated era ({\it i.e.} for temperatures
greater than the temperature at matter-radiation equality $T_{\rm eq} \approx 1~{\rm eV}$)
as long as the helicity is maximal \cite{Kahniashvili:2012uj}. 
So, from the relations in Eqs.~(\ref{calE}) and (\ref{helicity}), the energy density
scaling is ${\cal E} \propto a^{-4} \times a^{-2/3}$ in the radiation dominated era and
${\cal E}\propto a^{-4}$ in the matter dominated era. With these scalings, and with the cosmic 
cooling rate $T \propto a^{-1}$ and the temperature at big bang nucleosynthesis (BBN)
$T_{\rm BBN} \sim 1~{\rm MeV}$, we get 
$\Omega_{B\gamma {\rm BBN}} \sim 10^{-4} (1~{\rm kpc}/l_{d0} )^2$. 
Requiring $\Omega_{B\gamma {\rm BBN}} \lesssim 1$, this means that the 
magnetic dissipation scale today (also the coherence scale) is observationally constrained to be 
larger than $\sim 10~{\rm pc}$.


Spectral distortions of the cosmic microwave background (CMB) also provide a means to probe small 
scale magnetic fields for cosmological redshift $z$ between $10^3$ and 
$10^6$ \cite{Jedamzik:1999bm,Miyamoto:2013oua,Kunze:2013uja,Wagstaff:2015jaa}.
As of now the bounds from COBE/FIRAS measurements of the CMB spectrum are not competitive
with the BBN bound. Proposed experiments, such as PIXIE, can change this situation and be able
to detect CMB $\mu-$distortions for $l_{d0} \sim 1~{\rm kpc}$ (see Figs.~2 and 3 of 
Ref.~\cite{Wagstaff:2015jaa}). Small scale magnetic fields may also leave an imprint on the
CMB anisotropies through non-linear 
effects \cite{Sethi:2004pe,Jedamzik:2013gua,Kunze:2014eka,Chluba:2015lpa}.

The estimate in Eq.~(\ref{EGauss}) shows that intergalactic magnetic fields that are indicated by gamma 
ray observations may be of $\sim 3\times 10^{-10}~{\rm G}$ strength on $1~{\rm kpc}$ scales. During 
structure formation,
the field would get compressed within galaxies by a factor $(\rho_{\rm gal}/\rho_{\rm c})^{1/3}$, where
$\rho_{\rm gal} \approx 10^{-24}~{\rm gm}/{\rm cm}^3$ is the baryonic density in the galactic disk and
$\rho_{\rm c}\approx 10^{-31}~{\rm gm}/{\rm cm}^3$ is the cosmic baryon density.
If we assume flux freezing during structure formation, the magnetic field strength will increase
by $(\rho_{\rm gal}/\rho_{\rm c})^{2/3} \approx 10^5$ and the coherence scale will decrease by
$(\rho_{\rm c}/\rho_{\rm gal})^{1/3} \approx 10^{-2}$. With these numbers, and $l_{d0} = 1~{\rm kpc}$,
a galaxy would inherit a magnetic field with strength
$\sim 3\times 10^{-5}~{\rm G}$ and coherence $\sim 10~{\rm pc}$. A somewhat larger value of
$l_{d0} \sim 10~{\rm kpc}$ would lead to estimates that are closer to observations of 
the random component of the Milky Way magnetic field, $4-6~{\rm \mu G}$ on 
$10-100~{\rm pc}$ \cite{1993MNRAS.262..953O}.
This conclusion is in line with that of Ref.~\cite{Banerjee:2003xk} where the authors
argue that magnetic fields in galaxy clusters may arise directly from the intergalactic magnetic
field.

%

We now turn to the helicity of the magnetic field, a quantity that is parity (P) odd and also odd under
combined charge and parity (CP) transformations. Hence observed non-zero magnetic helicity indicates a 
period of CP violation in the early universe, as is also necessary for the generation of the observed 
cosmic matter-antimatter asymmetry. Thus it is natural to compare the 
observed magnetic helicity 
to the cosmic baryon number density, $n_{b0} \approx 10^{-7}~{\rm cm}^{-3}$,
\begin{equation}
\eta_{Bb0} \equiv \frac{H_0}{n_{b0}} \sim 2\times 10^{24} ~ \left (\frac{1~{\rm kpc}}{l_d} \right )
\label{etaBb}
\end{equation}
This estimate raises a challenge for fundamental physics -- what processes can generate such a large
helicity to baryon number ratio? 

The simplest particle physics based scenarios of magnetogenesis are based on the evidence that 
a baryon number changing process via an electroweak sphaleron \cite{Manton:1983nd}
also produces magnetic fields with $\sim 10^2$ helicity \cite{Copi:2008he,Chu:2011tx}. Then the 
magnetic helicity is proportional to the baryon number and we get $\eta_{Bb0} \sim 10^2$ 
\cite{Cornwall:1997ms,Vachaspati:2001nb}. Even in 
the unbroken phase of the electroweak model, where the electroweak sphaleron solution
does not exist {\it per se}, we expect gauge field production to occur during changes
of Chern-Simons number which is necessary for baryon number violation.

A more realistic view of the production of cosmic matter asymmetry is that baryon number violating 
processes occur so as to produce both baryons and antibaryons but with a slight excess of baryon 
production. In terms of magnetic fields this means that both left- and right-handed helical fields 
are produced but with a slight excess of left-handed helicity that is given by the fundamental CP 
violation \cite{Copi:2008he}. 
%
%
Within the context of baryogenesis in the standard model, CP violation is extremely weak 
\cite{Cline:2006ts} and the total helicity is tiny compared to the energy density in the magnetic 
field \cite{Copi:2008he}. If we assume energy equipartition,
the energy density in magnetic fields will be comparable to that in other forms of 
radiation \cite{Vachaspati:1991nm}.

The above description shows that the energy density in magnetic fields may be much larger than
that implied by magnetic helicity alone. However, the problem we are encountering based on observation,
is that the initial magnetic {\it helicity} also needs to be much larger (see Eq.~(\ref{etaBb})). Is there some 
dynamics beyond standard MHD that could potentially increase the magnetic helicity and saturate the 
maximal helicity condition in the early universe?

A simple possibility is to look for a mechanism that selectively amplifies one handedness of
the magnetic field. Then, if we start with a magnetic field, even with zero net helicity, the dynamics 
will amplify one of the two helicities, increase the magnetic field energy density, and also
saturate the helicity at its maximal value. This has been the focus of earlier studies of the ``chiral 
magnetic effect'' \cite{Vilenkin:1980fu}, in which a magnetic field induces an electric current 
${\bm j} \propto {\bm B}$, which results in the amplification of certain Fourier modes of
only one handedness \cite{Joyce:1997uy}. More importantly for us, however, the chiral magnetic effect 
also selectively {\it dissipates} one handedness of the magnetic field (see, for example, \cite{Tashiro:2012mf}). 
Thus, if baryon number violating interactions (or other dynamics) produce a large but non-helical 
magnetic field, the chiral magnetic effect can dissipate one of the two helicities -- the handedness 
being determined by the sign of the chiral imbalance -- and reduce the
magnetic field energy by half, and saturate the helicity at its maximal value.

More quantitatively, ignoring the plasma velocity field, the equations satisfied 
by the difference of the two helical amplitudes of the magnetic field Fourier
modes, $\Delta B \equiv |B^+(k)| - |B^-(k)|$, is given by (see Eq.~(60) in \cite{Tashiro:2012mf}),
\begin{equation}
\partial_\eta \Delta B =  +  \frac{k_p}{\sigma} (|B^+| + |B^-|) + \mathcal{O} (\Delta B)
\end{equation}
where $\eta$ is the conformal time, $\sigma$ is the electrical conductivity of the plasma, 
$k_p = {e^2\Delta\mu}/{2\pi^2}$, and the chemical potential 
$\Delta\mu = \mu_L-\mu_R$ is a measure of  the chiral imbalance of the medium. Thus
$\Delta B$ grows in proportion to the summed amplitudes of the two helicities of the magnetic 
field and the field tends to become maximally helical on a time scale set by the chirality of the medium.
(A chiral imbalance might arise naturally above the electroweak scale since the weak 
interactions distinguish between left- and right-handed particles at a fundamental
level.) Once the field becomes maximally
helical, it stays maximally helical. 
The precise dynamics, however, needs further investigation
since the analysis outlined above ignores the plasma velocity field.
The joint evolution of the magnetic field and the plasma velocity is essential to 
see effects such as the inverse cascade of helical fields. In a chiral medium, it is also
necessary to co-evolve the chemical potentials. The joint evolution of homogeneous chiral
imbalance has started to receive attention \cite{Schober2015} but even the
equations necessary to describe dynamics with spatially varying chirality have
not yet been established (recent attempts can be found in \cite{Boyarsky:2015faa,Gorbar:2016qfh}).

The main point of this paper is that current observational evidence for intergalactic
magnetic fields has profound implications for fundamental interactions. The observed 
magnetic fields must have originated in the early universe since they are seen in
voids and are helical. If we uncover a red spectrum of the magnetic field,
we would know that they were generated by an acausal mechanism. The magnetic fields 
would then provide valuable information about the earliest moments of the universe. If the 
spectrum turns out to be blue, the properties of the magnetic field will give us important 
clues about particle physics beyond the standard model.
The observation of magnetic helicity implies a strong role for fundamental
CP violation in the early universe. Since
helical magnetic fields are closely connected with baryon number violating processes,
the observation of helical magnetic fields can inform us about matter-genesis. 
But baryogenesis by itself is insufficient to explain the large helicity that is indicated
by observations. We have suggested that there may be a role for the chiral magnetic
effect to drive magnetic helicity to its maximal value. Then the standard model must be
extended to allow for successful baryogenesis {\it and} the chiral magnetic effect should
play a role in cosmology. This
would have implications for particle physics close to the electroweak scale and may 
perhaps also be testable at the LHC or future accelerator experiments. 
Future observations ({\it e.g.} by the Cherenkov Telescope Array \cite{Meyer:2016xvq})
will further sharpen the case for intergalactic magnetic fields and allow
for more precise measurements of the power spectra. 


\acknowledgements

I thank Jens Chluba, Francesc Ferrer, Tina Kahniashvili, Andrew Long, Levon Pogosian and 
Alex Vilenkin for comments and 
suggestions. I am grateful to the Institute for Advanced Study, Princeton for hospitality while 
this work was being done. This work was supported by the U.S. Department of Energy, Office 
of High Energy Physics, under Award No. DE-SC0013605 at ASU.

\end{document}